\documentstyle[aps,epsf,amssymb,amsfonts,graphics,prl,multicol]{revtex}

\draft

\begin{document}
\title{A Comparison of Entanglement Measures}
\author{Jens Eisert$^{1}$ and Martin B. Plenio$^{2}$} 
\address{
\protect\small\em $^{1}$ Institut f{\"u}r Physik, Universit{\"a}t
Potsdam, Germany\\
\protect\small\em $^{2}$ Blackett Laboratory, Imperial
College, Prince Consort Road, London SW7 2BZ, U.K.}

\maketitle

\date{\today}

\begin{abstract}
We compare the entanglement of formation with a measure defined as the
modulus of the negative eigenvalue of the partial transpose. In particular 
we investigate whether both measures give the same ordering of density
operators with respect to the amount of entanglement.
\end{abstract}


\begin{multicols}{2}
\narrowtext

\section{Introduction}
Entanglement is a key property that makes quantum information theory
different from its classical counterpart \cite{Plenio98c}. Maximally entangled states, 
for example, are the basis of quantum state teleportation which has 
recently been demonstrated \cite{Bennett93,DeMartini98,Zeilinger98}. 
Under realistic conditions, however, one will only be able to generate
partially entangled mixed states. It is then of interest to be able 
quantify the amount of entanglement in such states. Some measures
of entanglement for mixed states have been suggested recently
\cite{Bennett96c,Plenio97a,Plenio97b,Plenio98a}. They are 
useful, for example, as upper bounds to entanglement purification protocols
\cite{Bennett96a,Bennett96b,Gisin96,Horodecki97b,Murao98} and to the 
quantum channel capacity of certain quantum communication channels
\cite{Bennett96c,Plenio98a,Rains97}.

Unfortunately entanglement measures for mixed states (which are relevant 
in the presence of noise) are usually quite hard to calculate
analytically,
although an analytic expression for the entanglement of formation of
two spin-$1/2$ particles is now known \cite{Wootters98}. The general case
remains  
unsolved. For some problems, however, it is not so important to know
the exact amount of entanglement (a quantity that is not unique
anyway). It would be completely sufficient if one knew which state of 
a family of states has the most entanglement. To answer this question it 
would be sufficient to find a (hopefully as simple as possible) quantity 
that preserves the ordering of density operators with respect to
entanglement, 
i.e. that for two measures $E_1$ and $E_2$ and any two density operators 
$\rho_1$ and $\rho_2$  we have that $E_1(\rho_1)>(<)E_1(\rho_2)$ is
equivalent
to $E_2(\rho_1)>(<)E_2(\rho_2)$.

In this paper we will compare the entanglement of formation for two
spin-$1/2$ particles, for which a closed analytical form is known, 
with a quantity which was proposed in \cite{Horodecki96} as a way to 
quantify the degree of entanglement of a mixed state. The basis of this 
`measure of entanglement' is the Peres-Horodecki criterion for the 
separability of bipartite systems \cite{Horodecki96,Peres96}. Given a
state of, for example, two spin-$1/2$ systems one calculates the partial
transpose of the density operator. The state is separable exactly if the 
partial transpose is again a positive operator. If, however, one of the
eigenvalues of the partial transpose is negative then the state is
entangled. 
One can now imagine that the amount of entanglement is quantified by the 
modulus of this negative eigenvalue, i.e. the larger it is, the larger the
entanglement of the state. 

It is important to check whether this measure indeed preserves the ordering 
of density operators as it has been used for this purpose in some
publications.

In section II we will explain the entanglement of formation and
subsequently we summarize some properties of
the negative eigenvalue measure of entanglement. Finally, we present 
in section III both a numerical and an analytical comparison of the two 
measures of entanglement with respect to the ordering of density operators 
induced by them. In section IV we sum up the results of this paper.

\section{Entanglement measures}
In this section we briefly describe some entanglement measures in particular
the entanglement of formation and the negative eigenvalue measure of entanglement.

There are not very many good measures of entanglement. One example is the 
relative entropy of entanglement \cite{Plenio97a,Plenio97b,Plenio98a}. In fact, 
it gives rise to the most restrictive upper bound on the channel capacity of 
the depolarizing channel \cite{Plenio98a,Rains97}.
Unfortunately, even for two spin-$1/2$ particles no general analytical
expression has been found for it so far although many special cases can be
solved analytically.

A second measure of entanglement and actually the first measure of
entanglement that has been proposed for mixed states is the entanglement 
of formation \cite{Bennett96c}. It basically describes the amount of
entanglement that needs to be shared previously in order to be able to create a
particular ensemble in a given state $\rho$ by local operations. Mathematically, 
this means that we find that pure state ensemble that realizes the state $\rho$
and which has the smallest amount of entanglement, i.e.,
\begin{equation}
        E_F(\rho) = \min_{\rho = \sum_i p_i |\psi_i\rangle\langle \psi_i|}
                      \sum_i p_i E_{vN}(|\psi_i\rangle\langle \psi_i|)
\;\; ,
        \label{formation}
\end{equation}
where $\{|\psi_i\rangle\}$ is a set of not necessarily orthogonal pure
states. The entanglement of formation is known to be larger than the
relative
entropy of entanglement which proves that in general quantum state
purification
methods cannot recover all the entanglement that has been invested in the
creation 
of the quantum state \cite{Plenio98a}. 

A nice feature of the entanglement of formation is the fact that it can be
solved analytically for a system of two spin-$1/2$ particles
\cite{Wootters98}. This allows for fast numerical studies as the
cumbersome 
minimization Eq. (\ref{formation}) can be avoided.
 
The entanglement of formation can be expressed in terms of the function
\begin{equation}
        {\cal E}(C) = h\left( \frac{1 + \sqrt{(1-C^2)}}{2} \right)
        \label{ent1}
\end{equation}  
where 
\begin{equation}
        h(x) = -x \log_2 x - (1-x) \log_2(1-x) \;\; .
        \label{entropy}
\end{equation}
For a density operator $\rho$ one defines the spin flipped state
\begin{equation}
        {\tilde \rho} = (\rho_y \otimes \rho_y) \rho^* (\rho_y \otimes
\rho_y)
        \label{flip}
\end{equation}
where the ${}^*$ denotes the complex conjugate in the standard
computational
basis $\vert 00\rangle,\vert 01\rangle,\vert 10\rangle,\vert 11\rangle$.
One then finds the entanglement of formation to be 
\begin{equation}
        E_F(\rho) = {\cal E}(C(\rho))
        \label{formation1}
\end{equation}
where the so called concurrence is defined by 
\begin{equation}
        C(\rho) = \max \{0,\lambda_1-\lambda_2-\lambda_3-\lambda_4\}.
\end{equation}
Here, $\lambda_1,...,\lambda_4$ are the eigenvalues, in decreasing order,
of the Hermitean
matrix $R= \sqrt{\sqrt{\rho}{\tilde \rho}\sqrt{\rho}}$. 
For properties of this
measure of entanglement the reader should consult the literature
\cite{Bennett96c,Wootters98}. It is interesting to note that
since ${\cal E}$ as a function of the concurrence $C$
is a strictly monotonous function and maps the interval $[0,1]$ on
$[0,1]$ $C$ can in fact also be regarded as a measure for
entanglement. 

We will now consider the negative eigenvalue of the partial transpose of a 
density operator as a measure of entanglement. In the next section we
will then compare it to the entanglement of formation. 

For two spin-$1/2$ 
particles (which form the two systems $A$ and $B$)
any disentangled state $\rho$ can be written as the convex sum 
of product states
\begin{equation}
        \rho = \sum_i p_i \rho_i^A \otimes \rho_i^B\;\; .
        \label{separa}
\end{equation}
States which permit a representation of the form Eq. (\ref{separa}) are
also called separable. For two spin-$1/2$ particles there is a simple
criterion to decide whether a given state is separable or not
\cite{Peres96,Horodecki96}. One calculates the partial transpose
$\rho^{T_B}$ 
of the density operator $\rho$ in the computational basis. This means that 
we transpose only one subsystem, either
subsystem $A$ or $B$. If the resulting matrix is positive semidefinite 
then the density operator $\rho$ is separable; otherwise it is not. 
Therefore the partial transpose of an entangled state has a negative 
eigenvalue and the idea of the negative eigenvalue measure is to use
the modulus of the negative eigenvalue to quantify the entanglement 
of the state $\rho$. In a mathematical form this reads as
\begin{equation}
        E_N(\rho) =
|\min
\{0,\lambda_1^{T_B},\lambda_2^{T_B},\lambda_3^{T_B},\lambda_4^{T_B}\}|
        \label{Horo}
\end{equation}
where the $\lambda_i^{T_B}$ are the eigenvalues of the partial transpose
$\rho^{T_B}$.
However, we do not know whether this way of quantifying the entanglement 
constitutes a proper measure of entanglement. Therefore we will investigate 
in the next section numerically whether the entanglement of formation and 
the negative eigenvalue measure are compatible from a different point of view. 

\section{Ordering induced by entanglement measures}

We would expect that any two `good' entanglement measures should generate
the same ordering of the density operators. This means that for two
entanglement 
measures $E_1$ and 
$E_2$ and any two density operators $\rho_1$ and $\rho_2$ we have that 
\begin{equation}
        E_1(\rho_1) > E_1(\rho_2) \Leftrightarrow E_2(\rho_1) >
E_2(\rho_2)\;\; .
        \label{consist}
\end{equation}
Why do we expect this relation to be true? If in one measure of
entanglement
$\rho_1$ contains more entanglement than $\rho_2$ then we would expect 
that a quantum state purification method would generate more singlets from
an ensemble in state $\rho_1$ than ensemble $\rho_2$. If $E_2$ is also a
measure of entanglement then we would expect that $\rho_2$ would yield
more singlets than $\rho_1$. While this reasoning is not strict, it
nevertheless indicates that Eq. (\ref{consist}) should be true for 
two `good' measures of entanglement.

In the next two subsections we will now check whether Eq. (\ref{consist})
is satisfied for the entanglement of formation $E_F$ and the negative
eigenvalue measure of entanglement $E_N$. 

\subsection{An analytical comparison}
For some classes of density operators we can easily check analytically 
whether the relation Eq. (\ref{consist}) is true. For pure states it is
sufficient to consider states of the form 
$|\psi\rangle = \alpha |00\rangle + \beta |11 \rangle$ 
as it follows from the Schmidt decomposition \cite{Schmidt}. The
entanglement of formation $E_F$ then reduces to the von Neumann entropy of 
entanglement while the negative eigenvalue measure yields
\begin{equation}
        E_N(|\psi\rangle\langle \psi|) = \alpha
\beta=\alpha\sqrt{1-\alpha^2}.
        \label{pure}
\end{equation}
Both measures decrease monotonically with $\alpha$ so that for pure states
Eq. (\ref{consist}) is satisfied. It should be mentioned that 
for pure states $|\psi\rangle\langle\psi|$
with $|\psi\rangle = \alpha |00\rangle + \beta |11 \rangle$
the concurrence $C$ is given by
\begin{equation}
	C(|\psi\rangle\langle\psi|)= 2\alpha\sqrt{1-\alpha^2};
\end{equation}
hence, also for arbitrary pure states $\rho=|\psi\rangle\langle\psi|$
the negative eigenvalue 
measure and the concurrence are connected by the simple equation
\begin{equation}\label{Connection}
	C(\rho)=2E_N(\rho).
\end{equation}
Werner states are defined as
\begin{equation}
        \rho_F = \frac{4F-1}{3}|\psi^{-1}\rangle\langle \psi^{-1}| +
\frac{1-F}{3} {\bf 1},
        \label{Werner}
\end{equation}
where $|\psi\rangle = (|01\rangle - |10\rangle)/\sqrt{2}$ is the singlet
state and 
$F\in [1/4,1]$. 
The concurrence of Werner states is given by $C(\rho_F)=2F-1$
for $F \ge 1/2$ (otherwise $\rho_F$ is a separable state)
while the
entanglement of formation is found to be
\begin{equation}
        E_F(\rho_F) = - \mu \log_2 \mu  - (1-\mu) \log_2 (1- \mu)
        \label{werfor}
\end{equation}
with $\mu = 1/2+\sqrt{F(1-F)}$.
The negative eigenvalue measure simply gives
\begin{equation}
        E_N(\rho_F) = F - \frac{1}{2},
        \label{werneg}
\end{equation}
which means that Eq. (\ref{Connection}) is also valid for Werner states.
Again, both measures decrease monotonically with $F$ and therefore Eq.
(\ref{consist}) is satisfied. 

Moreover, no counterexample to Eq. (\ref{consist}) can be
constructed from pure states and Werner states, i.e. no two
density operators $\rho_1$ and $\rho_2$ can be found which violate
Eq. (\ref{consist}) when $\rho_1$ is assumed to be a pure state and
$\rho_2$ corresponds to a Werner state. 
Because of the
monotony of ${\cal E}$ in Eq. (\ref{ent1}) this can already 
be seen when comparing the concurrences of $\rho_1$ and $\rho_2$
with the negative eigenvalue measure of the respective states:
from Eq. (\ref{Connection}) it then immediately follows that
\begin{equation}
        C(\rho_1) > C(\rho_2) \wedge E_N(\rho_1) < E_N(\rho_2)
\end{equation}
cannot be fulfilled for any value of $F$.

One may therefore suspect that Eq.  (\ref{consist}) also holds  
for arbitrary mixed states.
To tackle the question whether  Eq.  (\ref{consist})
is generally satisfied and hence the two entanglement
measures induce the same ordering we have 
employed Monte Carlo simulations. If Eq.  (\ref{consist}) is 
violated at all, it is - roughly speaking - 
also of interest  to investigate
how `badly' it is violated after all.
In the following subsection we present the results of the numerical
test that we have performed.

\subsection{A numerical comparison}
In order to test the ordering induced by the
two entanglement measures under consideration 
we have numerically generated a million pairs of 
random density matrices with respect to a certain well defined
distribution. In the case of two spin-$1/2$ particles, which is the case
of interest in this paper, one has to generate random density operators -
i.e.  linear, self adjoint, positive semidefinite operators $\rho$
of finite trace $(=1)$ - acting on the
four dimensional Hilbert space isomorphic to 
${\Bbb{C}}^2\otimes {\Bbb{C}}^2 $ .
One can represent $\rho$ by the decomposition
\begin{equation}
	\rho=\sum_{i=1}^4 p_i |\psi_i\rangle\langle\psi_i|
\end{equation}
with pairwise orthogonal projections,
where $p_i\geq0$ and $\sum_i p_i=1$. This in turn
corresponds to a $4\times4$ matrix of the form
\begin{equation}\label{Construction}
	\rho= U D U^\dagger,
\end{equation}
where the diagonal matrix $D$ is defined by $D_{ij}=\delta_{ij}p_i$.

A plausible choice for the ensemble of unitary $4\times4$
matrices which can be used to construct 
density matrices according to Eq. (\ref{Construction}) 
is the one with the normalized 
Haar measure 
on the group of unitary matrices $U(4)$
which is called the circular unitary ensemble \cite{Remark}.
It has been shown in \cite{UnitaryMatrices}
that random unitary matrices representative for the
circular unitary ensemble can be obtained as follows: 
Let us set
\begin{eqnarray}
	U&&=U^{(1,2)}(\phi_{12},\psi_{12},\chi_{12})\\
	&&\times
	U^{(2,3)}(\phi_{23},\psi_{23},0)
	U^{(1,3)}(\phi_{13},\psi_{13},\chi_{13}\nonumber)\\
	&&\times
	U^{(3,4)}(\phi_{34},\psi_{34},0)
	U^{(2,4)}(\phi_{24},\psi_{24},0)
	U^{(1,4)}(\phi_{14},\psi_{14},\chi_{14}),\nonumber
\end{eqnarray}
where the 
complex $4\times4$ matrices $U^{(i,j)}$, $i,j=1,...,4,$ 
with three real parameters $\phi$, $\chi$, and $\psi$ are given by
\begin{equation}
	U^{(i,j)}_{kl}(\phi,\psi,\chi)=\left\{
		\begin{array}{ll}
			1,\;\;\;&k=l,\;k\neq i,j,\\
			\sin\phi e^{i\chi},\;\;\; &k=i,\;l=j,\\
			\cos\phi e^{i\psi},\;&k=l=i,\\
			\cos\phi e^{-i\psi},\;&k=l=j,\\
			-\sin\phi e^{-i\chi},\;\;&k=j,\;l=i,\\
			0, &\text{otherwise}.\\
		\end{array}
	\right.
\end{equation}
We now take $\psi_{ij}$ and $\chi_{ij}$ to be independent
random variables with a uniform distribution in the interval $[0,2\pi)$,
and for a given random number $\xi$ distributed uniformly in the
interval $[0,1)$ 
we choose $\phi_{ij}$ 
to be $\arcsin(\xi^{1/(2i)})$ for $i=1,...,3$.
Matrices generated in this way
are then random unitary matrices of the wanted type (except for a random
phase which is of no significance for our purposes).
Finally, a set of random density 
matrices can be obtained
by appropriately choosing the diagonal matrices
$D$. More precisely, the random vector with entries $p_1,...,p_4$ 
should be uniformly distributed on the manifold defined by $\sum_i p_i=1$.
According to 
\cite{Volume} this can be achieved, e.g.,
by setting 
$p_1=1-\xi_1^{1/3}$,
$p_2=(1-\xi_2^{1/2})(1-p_1)$, 
$p_3=(1-\xi_3)(1-p_1-p_2)$, and
$p_4=1-p_1-p_2-p_3$, where again,
$\xi_1$, $\xi_2$, and $\xi_3$ are random numbers drawn with respect to a  
uniform distribution in the interval $[0,1)$. 

\begin{figure}
$\Delta E_F(\rho_1,\rho_2)$
\vspace{-0.5cm}

\centerline{
\rotatebox{-90}{
	\epsfxsize=6.5cm
	\epsfbox{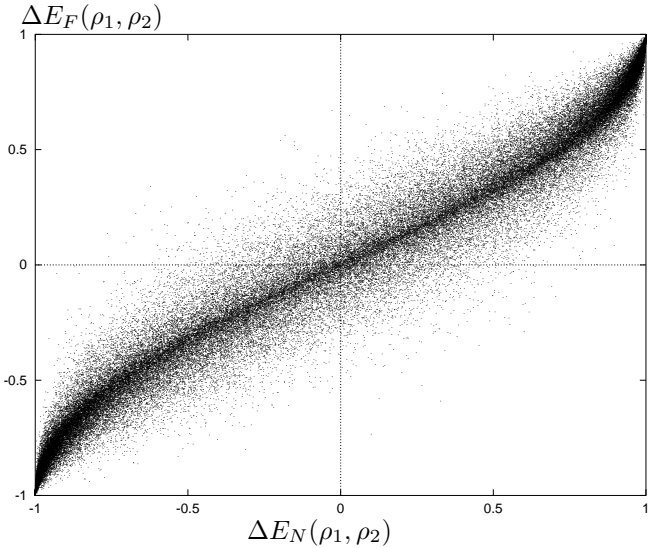}
}
}
\centerline{$\Delta E_N(\rho_1,\rho_2)$}

\smallskip

\caption{$\Delta E_F(\rho_1,\rho_2)$ versus $\Delta E_N(\rho_1,\rho_2)$ 
for $10^4$ pairs of entangled states $(\rho_1,\rho_2)$.
Pairs of states with $\Delta E_F(\rho_1,\rho_2)\Delta
E_N(\rho_1,\rho_2)<0$
are represented by dots in the second and fourth quadrant.}
\label{Fig1}
\end{figure}

To investigate whether Eq.  (\ref{consist})
is satisfied we now draw random density matrices from the
previously described ensemble and
calculate the eigenvalues of the partial transposes
and the eigenvalues of the products of the respective 
density operators with the spin flipped states.
After discarding those
density matrices which correspond to separable
states we then determine values for $E_N$ and $E_F$
as described in section II. Finally, we
check for pairs $(\rho_1,\rho_2)$ 
whether the sign of
\begin{equation}
	\Delta E_F(\rho_1,\rho_2)=
\frac{E_F(\rho_1)-E_F(\rho_2)}{E_F(\rho_1)+E_F(\rho_2)}
\end{equation}
and
\begin{equation}
	\Delta E_N(\rho_1,\rho_2)=
  \frac{E_N(\rho_1)-E_N(\rho_2)}{E_N(\rho_1)+E_N(\rho_2)}
\end{equation}
is identical. Fig. \ref{Fig1} shows a diagram in which 
$\Delta E_F(\rho_1,\rho_2)$ is plotted versus 
$\Delta E_N(\rho_1,\rho_2)$ for $10^4$ pairs $(\rho_1, \rho_2)$ of
random density matrices. Although there is obviously a certain
correlation between $\Delta E_F$ and $\Delta E_N$,
there are dots in the second and
the fourth quadrant of the diagram
which are associated with pairs of states which do not satisfy
Eq.  (\ref{consist}). One can therefore conclude that in the case 
of two spin-$1/2$ systems this relation is not satisfied
for arbitrary mixed states, and hence, the ordering induced
by the entanglement of formation $E_F$ and the one induced by 
the negative eigenvalue measure $E_N$ is not the same.

It is interesting to note that most of the randomly drawn density
operators
do not correspond to entangled but to separable states; so more density 
matrices have to be discarded than can be kept for the investigation.
In fact, from the numerical simulations we can estimate the probability
$P_E$ that
a created mixed state is entangled as
$ P_E\approx 0.365\pm 0.001$, which
is in complete agreement with the analytical and
numerical findings presented in \cite{Volume}. 

From the Monte Carlo simulation we can also give a rather
accurate estimate of how likely it is to find a pair
of states which violates Eq. (\ref{consist}),
given that both states are entangled. From the
relative frequency of a violation 
for a million pairs of density matrices 
we obtain
for the estimate of this probability $P_V$
\begin{equation}
	P_V\approx0.047\pm 0.001.
\end{equation}
\begin{figure}
$E_N$
\vspace{-0.5cm}

\centerline{
\rotatebox{-90}{
	\epsfxsize=6.5cm
	\epsfbox{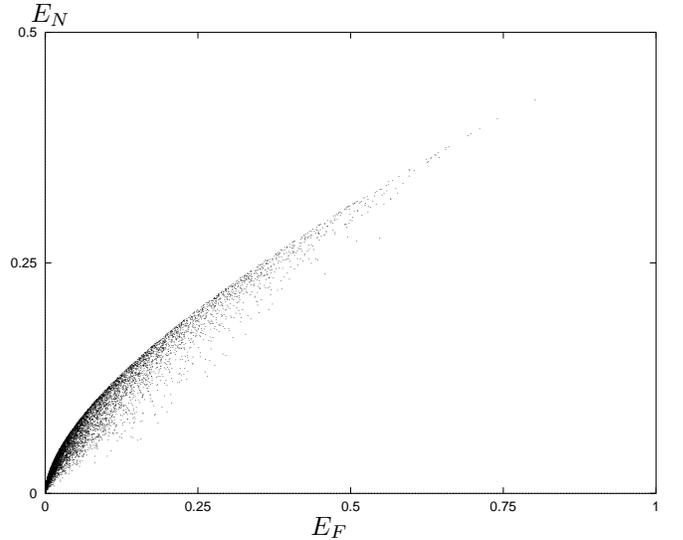}
}
}\centerline{$E_F$}

\caption{Distribution of states: $E_N$ versus $E_F$.}
\label{Fig2}
\end{figure}
Now that we know that the ordering is different it is
interesting to see to what extent the values of $E_N(\rho)$
and $E_F(\rho)$ differ for a given state $\rho$,
or how these random states are distributed in a diagram
where $E_F$ is plotted versus $E_N$. Fig. \ref{Fig2} shows such a plot.
Since the ordering induced by $E_F$ and $E_N$ is not the
same, there is no strictly monotonous function $f:[0,1]\rightarrow[0,1/2]$ such that
$E_N(\rho)=f(E_F(\rho))$ for all states $\rho$. This is also obvious from Fig. \ref{Fig2}.
In Fig. \ref{Fig3} again the distribution
of states is shown, but this time 
the negative eigenvalue measure is plotted versus the
concurrence $C$.
We can see that most of the dots are located close
to the diagonal connecting $(0,0)$ and $(1/2,1)$;
this diagonal corresponds to states satisfying 
Eq. (\ref{Connection}). Note that the numerical
simulation also strongly suggests that 
for a state $\rho$ with a certain value of $C(\rho)$
the upper bound for the possible values of the negative
eigenvalue measure $E_N(\rho)$ is given by $C(\rho)/2$, that is, that 
in general $C(\rho)\geq2E_N(\rho)$ holds for any state $\rho$. 

Furthermore, since for pure states the ordering induced by the
two entanglement measures is the same but in general it is not, it 
is of interest
to investigate the dependence of the probability that
Eq. (\ref{consist}) is violated for a pair of entangled states
$(\rho_1,\rho_2)$ on the `purity of those states' with respect to 
a certain characterization of the purity.
In Fig. \ref{Fig4} we show the relative number of pairs $(\rho_1,\rho_2)$
of states which do not satisfy Eq. (\ref{consist}) versus
$S=S_1+S_2$ based on a million pairs
of entangled states. Here, $S_1$ and $S_2$
are the linear entropies 
\begin{equation}
	S_i=tr(\rho_i-\rho^2_i),\;\;i=1,2,
\end{equation}
of $\rho_1$ and $\rho_2$, respectively,
as commonly employed, e.g.,
in decoherence studies \cite{Zurek}. Obviously, 
pure states correspond to a vanishing linear entropy.
\begin{figure}
$E_N$
\vspace{-0.5cm}

\centerline{
\rotatebox{-90}{
	\epsfxsize=6.5cm
	\epsfbox{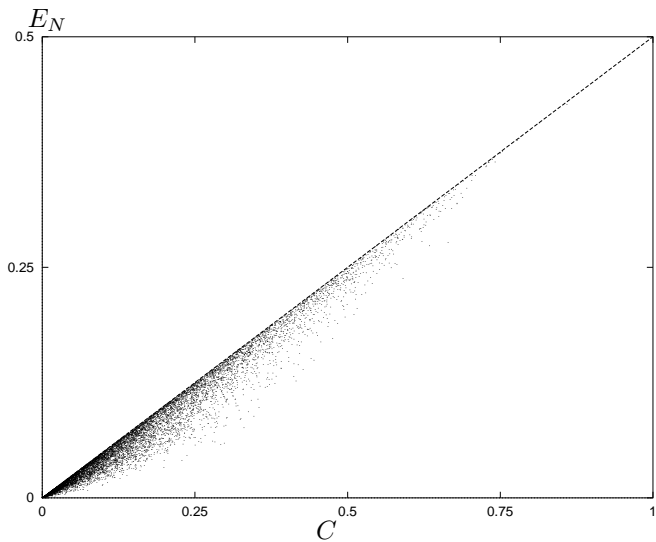}
}
}
\centerline{$C$}

\caption{Same as Fig. \protect\ref{Fig2} with $E_N$ versus $C$.}
\label{Fig3}
\end{figure}
\vspace*{-0.5cm}
We observe that the fraction of pairs $(\rho_1,\rho_2)$ of states  
with $\Delta E_N(\rho_1,\rho_2)
	\Delta E_F(\rho_1,\rho_2)<0$
increases monotonically with the sum of the
linear entropies of the respective states. This indicates
that 
the more mixed the two states are with respect to
the arithmetic mean of their linear entropies, the
larger is the probability that this pair violates 
Eq. (\ref{consist}).
Above a certain cut off no pairs of entangles states
can be found at all - the numerical data shown in Fig. \ref{Fig4}
give an estimate for the probability density 
of finding a pair of entangled states with a certain value of
$S$ (compare also \cite{Volume}).

It should finally be mentioned that in \cite{Volume} yet
another measure of entanglement incorporating the partial transpose
is proposed: For a given state $\rho$ the quantity 
$
	E=\sum_{i=1}^4 |\lambda_i^{T_B}|-1
$
is taken as a measure, where again, the $\lambda_i^{T_B}$
denote the eigenvalues of the partial transpose $\rho^{T_B}$ 
of $\rho$. While the actual value of $E$ for a given state
is of course different
from that of $E_N$,
from numerical simulations 
we come to the same conclusion that 
the ordering induced
by this measure is different from the one
induced by the entanglement of formation $E_F$.
\begin{figure} 
\vspace{-0.5cm}

\centerline{
\rotatebox{-90}{
	\epsfxsize=6.5cm
	\epsfbox{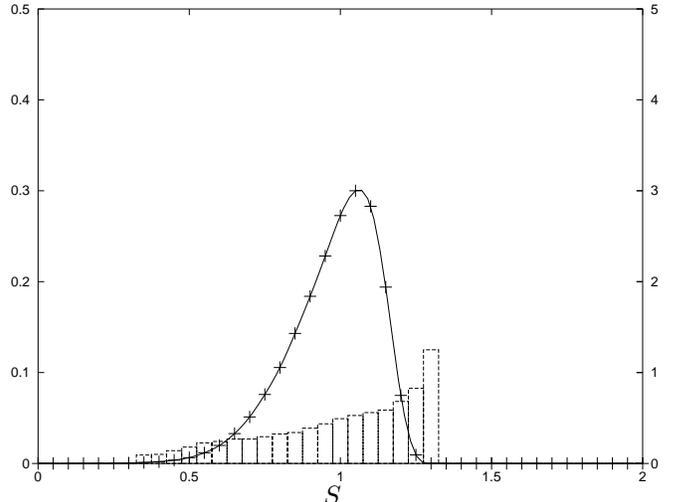}
}
}
\centerline{$S$}
\caption{Estimate for the probability density of finding a
pair of states $(\rho_1,\rho_2)$ with a certain value of 
$S=S_1+S_2$ ($+$). The solid line is a cubic spline
interpolation. Apparently, above a certain cut off
there are no pairs of entangled states any more. The histogram
shows the
relative number of pairs violating Eq. (\ref{consist})
compared to the total number of pairs of entangled states
for a given value of $S$. Note that the scaling is different
for both plots: the right axis belongs to the estimate 
for the probability density ($+$), the left one to the
relative number of violating pairs (histogram).}
\label{Fig4}
\end{figure}

\section{Summary}
We have compared the entanglement of formation with a
potential measure of entanglement that is given by the negative
eigenvalue of the partial transpose of the density operator 
of the system. In particular the ordering of density operators 
with respect to the amount of entanglement induced by the 
two measures has been compared both numerically and analytically.
We have shown that the negative eigenvalue measure does not
induce the same ordering as the entanglement of formation. Therefore
we do not expect it to be a `good' measure of entanglement. In particular
it cannot, in general, be used to determine the most entangled
state for a given family of density operators as it has been used previously.

\section{Acknowledgements}
The authors would like to thank Peter Knight, Vlatko Vedral, and
Martin Wilkens  
for discussions and useful hints. This work
was supported in part by the EPSRC, the European TMR Research Network
ERBFMRXCT960066 and the European TMR Research Network ERBFMRXCT960087.

\end{multicols}
\end{document}